\documentclass[12pt]{article}
\usepackage{epsfig}
\usepackage[utf8]{inputenc}
\usepackage[english,ukrainian]{babel}
\usepackage{rotating}

\begin{document}

\title{
MAVKA: PROGRAM OF STATISTICALLY OPTIMAL DETERMINATION OF PHENOMENOLOGICAL PARAMETERS  OF EXTREMA.
PARABOLIC SPLINE ALGORITHM AND ANALYSIS OF VARIABILITY OF THE SEMI-REGULAR STAR Z UMA
}
\author{Kateryna D.~Andrych$^{1,2}$,\\ Ivan L. Andronov$^{2}$, Lidia L.~Chinarova$^{3,2}$\\
$^{1}$ Department of Theoretical Physics and Astronomy,\\ Odessa I.I.Mechnikov National University, Ukraine,\\ {\tt katyaandrich@gmail.com}\\
$^{2}$ Department of Mathematics, Physics and Astronomy,\\
Odessa National Maritime University, Ukraine,\\
{\tt tt\_ari@ukr.net}\\
$^{3}$ Astronomical Observatory,\\
Odessa I.I.Mechnikov National University, Ukraine,\\
{\tt llchinarova@gmail.com}
}
\maketitle
\begin{abstract}
Advanced MAVKA software for the approximation of extrema observations is used to analyze the variability of the brightness of pulsating and eclipsing stars, but may be useful in analyzing signals of any nature.
A new algorithm using a parabolic (quadratic) spline is proposed.
In contrast to the traditional definition of a spline as a piecewise-defined function at fixed intervals, a spline is proposed to be divided into three intervals, but the positions of the boundaries between the intervals are additional parameters. The spline defect is 1, that is, the function and its first derivative are continuous and the second derivative can be discontinuous at the boundaries. Such a function is an enhancement of the "asymptotic parabola" (Marsakova and Andronov 1996).
The dependence of the fixed signal approximation accuracy on the location of the boundaries of the interval is considered. The parameter accuracy estimates using the least squares method and the bootstrap are compared. It is recommended to use the difference between the 0.975 and 0.025 percentiles (divided by 2*1.96) as the accuracy estimate of a given parameter in the bootstrap method.\\
The variability of the semi-regular pulsating star Z UMa is analyzed. The presence of multicomponent variability of an object, including, four periodic oscillations (188.88(3), 197.89(4) days and halves of both) and significant variability of the amplitudes and phases of individual oscillations are shown. Approximation using the parabolic spline is only slightly better than the asymptotic parabola, for our sampling of the complete interval. It is expectedly better for larger subintervals. The use of different complementary methods allows to get a statistically optimal phenomenological approximation.\\
\\[1ex]
\end{abstract}

\newpage

MAVKA: ПРОГРАМА СТАТИСТИЧНО ОПТИМАЛЬНОГО ВИЗНАЧЕННЯ ФЕНОМЕНОЛОГІЧНИХ ПАРАМЕТРІВ ЕКСТРЕМУМІВ.
АЛГОРИТМ ПАРАБОЛІЧНИХ СПЛАЙНІВ ТА АНАЛІЗ ЗМІН БЛИСКУ НАПІВПРАВИЛЬНОЇ ЗОРІ Z UMA

 К. Д.~Андрич$^{1,2}$, І. Л.~Андронов$^{1}$, Л. Л.~Чінарова$^{3,1}$\\
$^{1}$ Кафедра "Математика, фізика та астрономія",\\ Одеський національний морський університет\\
{\tt tt\_ari@ukr.net}\\
$^{2}$ Кафедра Теоретичної фізики та астрономії,\\ Одеський національний університет ім І.І.Мечникова\\
{\tt katyaandrich@gmail.com}\\
$^{3}$ Астрономічна обсерваторія,\\ Одеський національний університет ім І.І.Мечникова\\
{\tt llchinarova@gmail.com}

ABSTRACT.

Вдосконалено програмне забезпечення MAVKA для апроксимації рядів спостережень біля екстремуму. Воно використано для аналізу змінності блиску пульсуючих та затемнюваних зір, але може бути корисним при аналізі сигналів будь-якої природи.
Запропоновано новий алгоритм з використанням параболічного (квадратичного) сплайна.
На відміну від традиційного визначення сплайну, як кусково-заданої функції на фіксованих інтервалах, запропоновано сплайн із розділенням інтервалу на три підінтервали, але положення меж між підінтервалами є додатковими параметрами. Дефект сплайну дорівнює 1, тобто функція та її перша похідна неперервні, а друга може бути розривною на межах. Така функція є вдосконаленням "асимптотичної параболи" (Marsakova and Andronov 1996).
Розглянута залежність точності апроксимації фіксованого сигналу від розташування границь інтервалу. Проведено порівняння оцінок точності параметрів методом найменших квадратів та ''статистичного самотворення'' (bootstrap).\\
Проведено аналіз змінності напівправильної пульсуючої зорі Z UMa. Показано наявність багатокомпонентної змінності об'єкта, в тому числі, чотирьох періодичних коливань та суттєвої змінності амплітуд та фаз окремих коливань.


\section*{Вступ}

Досить часто з тих чи інших даних необхідно отримати феноменологічні параметри екстремуму, аби потім їх використовувати для подальшого фізичного аналізу процесу. Й зазвичай найнеобхідніші параметри – це момент та величина цього екстремуму, іноді, уся апроксимація, з відповідними оцінками точності (стандартними похибками).

Ці параметри використовують переважно для дослідження змін періодів затемнюваних подвійних систем та пульсуючих зір  [1-4]. Крейнер та ін. [5] опублікували 6-томну монографію, у якій зібрано 91\,798 моментів мінімумів 1\,138 затемнюваних систем. Чінарова та Андронов [6] визначили 6553 моменти екстремумів для 173 напівправильних зір та опублікували каталог цих та інших характеристик.

У подальшому, такі компіляції публікують на сайтах [7-9] та додатково у окремих статтях (напр. у журналі ``Open European Journal on Variable Stars'' [10])  по мірі накопичення даних. У Американській асоціації спостерігачів змінних зір AAVSO використовують скорочення ToM (`'time of minimum''). Втім, для пульсуючих зір визначають переважно моменти максимумів та не завжди моменти мінімумів, тому правильно було б розшифровувати скорочення ToM також і як (`'time of maximum'').

Історично, до одних з найперших методів відносять графічний метод Пікерінга, коли апроксимацію на міліметровці проводили ``від руки'', ''на око'', а потім проводили на різних рівнях горізонтальні лінії, визначали середини хорд та проводили через них пряму до перетину із апроксимацією. Із розвитком обчислювальної техніки (починаючи із арифмометрів), стали застосовувати апроксимацію параболою (чи, для асиметричних кривих, поліномом третього ступеня). Втім, найбільш популярним методом для (теоретично) симетричних кривих, став метод Кві та ван дер Воердена [11]. Оскільки відповідна тест-функція має багато локальних мінімумів, оцінка точності суттєво занижена, відповідно, вага такого значення неадекватно завищена.

 Андронов [12] використовував не лише поліноми ступеня, що забезпечує найкращу точність визначення моменту, а й поліноміальні сплайни. Апроксимація поліномами оптимального ступеня була реалізована різними авторами (напр., [6, 12, 13].

Сплайни із змінним ступенем були запропоновані [13], які виправляли проблему ''нефізичних'' хвиль на апроксимаціях поліномами. Огляди існуючих методів приведені [15-18].

Для зручності використання різних методів у одній програмі, було створено комп'ютерну програму MAVKA, у яку поступово додавалися різноманітні функції [19-21].

У цій статті, наведено порівняння двох базисних функцій: «асимптотична парабола» (запропонована [14]) та параболічний сплайн дефекту 1.
Обидва методи поділяють наведений інтервал поблизу екстремуму на три частини. Кожен з підінтервалів апроксимується власною функцією, що дозволяє зменшити похибки результатів та краще описати спостереження, врахувати асиметричність чи інші особливості.

Проведено аналіз впливу ширини інтервалу даних (лише екстремуму або й позаекстремальної частини) та асиметрії його меж на якість апроксимації та відповідні похибку визначених характеристик екстремуму. Для цього, обидва методи були застосовані до набору згенерованих даних з нормальним розподілом шуму.
Також  методи було проілюстровано застосуванням до реальних спостереженнях змінної пульсуючої зорі Z UMa із міжнародної бази даних AFOEV [22] із використанням методу статистичного самотворення (bootstrap), та проведено аналіз спостережень цієї зорі методами періодограмного та шкалограмного аналізу.

\section{Опис методів}

\subsection{Асимптотична парабола (Asymptotic Parabola, AP)}

	Цей метод був вперше запропонований у роботі В. І. Марсакової та І. Л. Андронова [14, 16] та показав свою ефективність при визначенні моментів екстремумів блиску змінних зір різних типів. Апроксимація складається з двох прямих ліній «асимптот», які з’єднані параболою між ними. При цьому, сама функція та її перша похідна неперервні. Попередній досвід використання цього методу показав, що цей метод є одним із найкращих для асиметричних максимумів пульсуючих зір.

Точки переходу від прямих на параболу $C_4$ та $C_5$.
\begin{equation}
x_c (t)=\left\{
\begin{array}{ll}
C_1+C_2 (-2 v-D)D+C_3 v,&~~ t < C_4,\\
C_1+C_2 v^2+C_3 v,&~~C_4\le t \le C_5,\\
C_1+C_2 (2 v-D)D+C_3 v,& ~~t > C_5,
\end{array}
\right.
\end{equation}
де $D=(C_5-C_4)/2$, $v=t-(C_5+C_4)/2$.

Із характерного вигляду функції зрозуміло, що дані, які будуть найефективнішими для такого наближення, будуть охоплювати не лише вузький інтервал біля екстремуму (який асимптотично буде співпадати із параболою), а й суттєву частину висхідної та нисхідної гілок кривої блиску, але не "суттєво закруглені" частини біля екстремумів протилежного виду (мінімумів, якщо визначаємо максимум, та навпаки). Вибір початкової та кінцевої точок зазвичай відбувається за допомогою програм OO [23], MCV [24] та також реалізовано у серії послідовно вдосконалених версій MAVKA [19-21].

Хоча метод був запропонований для визначення екстремумів, його можна використовувати й для визначення середнього моменту та характерного часу переходу з однієї асимптоти на іншу, навіть, при відсутності екстремуму. Наприклад, по діаграмі $"O-C"$ для моментів екстремумів (чи відповідних фаз) при порівняно різкій зміні періоду [25].

\subsection{Параболічний сплайн дефекту 1 (Parabolic Spline, PS)}

Поліноміальні сплайни є кусково-заданими функціями, де у кожному інтервалі аргументу (часу, фази обо ін.) задаються поліномом із своїм набором коефіцієнтів. Їх часто використовують для апроксимації даних. Вони характеризуються "порядком"  $n,$ або найвищим ступенем поліному, та "дефектом"  $k,$ який визначається, як найменша величина, для якої похідна порядку $(n-k)$ є неперервною в усіх точках, включаючи межі інтервалів [26-28].
	
Ця апроксимація також складається з трьох інтервалів, кожен з яких описується власною функцією (параболою). Точки переходу відповідають параметрам $C_6$ та $C_7$.

Параболічний сплайн відрізняється від попереднього методу двома додатковими параметрами, які описують параболічні складники у бічних інтервалах:
  	
\begin{equation}
x_c (t)=\left\{
\begin{array}{ll}
C_1+C_2 t+ C_3 t^2+C_4 (C_6-t)^2,&~~ t < C_6,\\
C_1+C_2 t+ C_3 t^2,&~~C_6\le t \le C_7,\\
C_1+C_2 t+ C_3 t^2+C_5 (t-C_7 )^2,& ~~t > C_7.
\end{array}
\right.
\end{equation}
Метод  дозволяє розширити ширину інтервалу апроксимації (й, таким чином, збільшити кількість точок, які використовуються) та отримати точність, яка відповідає результатам «асимптотичної параболи» на коротшому інтервалі.

Можливі ситуації, коли поділення інтервалу на три частини є неефективним. Наприклад, екстремум занадто гострий й центральна параболічна частина вироджується. Для такої ситуації, у програмі MAVKA [21] реалізовано зменшення кількості інтервалів (а, отже, й відповідних нелінійних параметрів). Таким чином, апроксимація «асимптотичною параболою» може складатися лише з двох асимптот, параболи та лише однієї асимптоти або взагалі лише параболи.

Так само, «параболічний сплайн», за занадто малих бічних інтервалів, автоматично замінюється на звичайну параболу.

В рамках реалізованого алгоритму, лінійні параметри були визначені методом найменших квадратів, а нелінійні – методом диференційних поправок.
Статистичні похибки параметрів екстремуму визначені за допомогою коваріаційної матриці похибок коефіцієнтів (напр. [30,16]).

\section{Порівняння асимптотичної параболи та параболічного сплайну дефекту 1 на штучних даних зі змінними межами}

Ми застосували обидва методи до набору з 25 штучних інтервалів. Як сигнал, було обрано синусоїду із одиничною півамплітудою. Шум був згенерований з нормальним розподілом випадкової величини із теоретичним середньоквадратичним значенням 0.1.
%? з максимальним відхиленням у 0.1 від ідеальної кривої.
Ми змінювали межі інтервалів між екстремальними випадками, наведеними на рис.1, для оцінки впливу ширини охоплення даних поза екстремумом на апроксимацію обома функціями.

При зменшенні ширини інтервалів та `'чистому'' сигналі (без шуму), очікується, що усі апроксимації будуть асимптотично сходитися до параболи (чи поліному більшого ступеня, найчастіше, кубічної параболи). Можливість апроксимації поліномом із автоматичним вибором ступеня полінома була реалізована на різних мовах програмування [6, 13,  19, 21], й також присутня й у програм MAVKA. 
Але, кількість точок зменшується, та шуми спостережень призводять до появи `'нефізичних'' хвиль на апроксимації, що нагадує ефект Гібса для тригонометричніх поліномів (напр. [35]). Це суттєво погіршує як якість апроксимації, так і може призвести до суттєвого зсуву визначеного моменту екстремуму.

Тому, для аналізу, мінімальна ширина інтервала була обмежена, як показано на рис. 1. Для меншої ширини, розділення інтервала апроксимації на 3 підінтервали стає формальним. Адже парабола може буть однаково представлена, як кубічний сплайн, при будь-якому діленні на підінтервали. Тест-функція стала, як функція двох змінних (границь підінтервалів), отже, матриця нормальних рівнянь у методі диференціальних поправок стає виродженою. 
В цьому випадку, апроксимація поводиться квадратною параболою, а не сплайном.
	
Результати на кожному тестовому інтервалі для відповідного значення меж були усереднені за відхиленнями від ідеального (синусоїди без шуму) значення та оцінками точності для подальшого аналізу.
%?
Як видно на Рис. 2, ''асимптотична парабола'' дуже чутлива до зміни меж апроксимації, на відміну від параболічного сплайну. Останній метод дозволяє проводити точну криву за наявності даних поза сусіднім екстремумом. Однак, коли інтервал звужується, обидва методи дають практично однакові результати, або в рамках відхилення випадково розподіленого шуму. Графіки у верхному рядку Рис. 2. відповідають симетричній зміні меж інтервалу, менша кількість точок відповідає вужчому інтервалу.

Для оцінки впливу асиметричності меж екстремуму, ми зафіксували ліву межу на одному положенні, а праву змінювали покроково. Крайні випадки меж ширини інтервалів для цього аналізу наведені у правому стовбчику Рис. 1. Як і в попередньому випадку, за занадто широкого інтервалу даних, асимптотична парабола програє параболічному сплайну. Однак далі, асимптотична парабола демонструє не гірші, а, часом, й кращі результати. Зі звуженням інтервалу, правий підінтервал апроксимацій вироджується, спричинюючи помилку в роботі та навіть виродження матриці нормальних рівнянь для апроксимації параболічним сплайном.

Метод асимптотичної параболи більш стійкий для таких змін. Для цього методу, реалізовано врахування випадків виродження якогось з підінтервалів. В таких ситуаціях, апроксимація відбувається лише лінійною асимптотою та параболою, або двома асимптотами.

Графіки у нижньому рядку Рис. 2 демонструють залежності параметрів екстремуму для асиметричної зміни меж інтервалу, менша кількість точок відповідає вужчому інтервалу.

Очікуваною є перевага апроксимації параболічним сплайном на широких інтервалах, оскільки кількість спостережень збільшується, а три параболи показують менше систематичне відхилення від сигналу, ніж асимптотична парабола.  Для невеликих інтервалів поблизу екстремума, перевагу має асимптотична парабола.

Загальна рекомендація при розмітці інтервалів для визначення екстремумів: вибирати (при можливості) приблизно однакові за тривалістью ділянки кривої блиску. Для асимптотичної параболи, брати якнайдовші відрізки висхідної та нисхідної гілок кривої блиску, де зміна блиску близька до лінійної.  З попереднього досвіду використання `''асиптотичних парабол'', блиск на границях інтервалу приблизно вдвічи ближчий до сусіднього екстремуму, ніж до апроксимованого.
Звісно, це залежить від конкретного виду кривої блиску. Для автоматизації процесу, й було запропоновано вибір між різними функціями для фіксованого інтервалу даних.
 
Незважаючи на якісний характер рекомендації, для спостережень із порівняно низьким значенням відношення оцінки точності до амплітуди (що характерно не лише для фотографічних та візуальних спостережень зір з великою амплітудою,  а й для величезної кількості малоамплітудних зір, відкритих із орбітальних обсерваторій), такий візуальний підхід (або різноманітні напівавтоматичні алгоритми) дає значення параметрів, які несуттєво відрізняються при невеликих змінах інтервалу. Але статистична оцінка точності порівняно із дуже широкими, чи, навпаки, вузькими, інтервалами краща на десяткі відсотків, або, навіть, у рази.

\section{Застосування обох методів до спостереженнь Z UMa}

Для подальшого аналізу, було використано 12578 спостережень Z UMa із міжнародної бази даних Французької асоціації спостережень змінних зір
(AFOEV) [22], отриманих у діапазоні юліанських дат 2451630 -- 2458026 (тривалість 17.5 років). Аналіз попередніх даних проведено [6].

Крива блиску пульсуючої зорі Z UMa була попередньо розмічена на інтервали поблизу екстремумів (як максимумів, так і мінімумів). Ми провели апроксимацію кожним методом окремо.

Ми аналізували точність моментів екстремуму, точність зоряної величини екстремуму та середньоквадратичне відхилення для обох методів відповідно до номеру інтервалу. Результати досить схожі, усі інтервали (згідно розмітки) не містили позаекстремальної частини. Іноді відповідну точність для асимптотичної параболи іноді все ж дещо кращі.

Також ми порівняли обидва методи, аби алгоритм програми MAVKA самостійно визначив найкращий метод для кожного конкретного інтервалу. Вибір кращого методу відбувається за меншою оцінкою точності моменту екстремуму.

В результаті, з 94 інтервалів, асимптотична парабола виявилась краща для 68 з них, а параболічний сплайн для 26. Медіана відношення оцінки точності моменту екстремуму для PS та AP трохи менша за одиницю (0.86). Але, для 34 інтервалів із 94, екстремум не попадає у центральний підінтервал апроксимації, і тому відбраковувався, хоча сама апроксимація була достатньо якісна.
З врахуванням цих відбракованих даних, медіана суттєво перевищує одиницю (2.36).

 Це пов'язано з декількома причинами. По-перше, сплайн менш стійкий до шуму в спостереженнях, ніж асимптотична парабола, яка іноді може визначати не глобальний екстремум на інтервалі, а локальну флуктуацію шуму. По-друге, на даний момент, апроксимація параболічним сплайном була реалізована за умови, що момент екстремуму має потрапляти в центральний підінтервал. Однак, можуть бути ситуації, коли апроксимація проведена досить точно, але екстремум знаходиться на одному з бічних підінтервалів, що призводить до вильоту алгоритму при спробі його розрахувати або завеликих та некоректних похибок визначених параметрів.

Медіана відношення середньоквадратичних відхилень спостережень від апроксимацій PS та AP становить 0.986, що лише несуттєво менше одиниці.

Асимптотична парабола є стійкішою до таких випадків, й дозволяє коректно визначити характеристики екстремуму для будь-якого екстремуму.
Тому ці методи не замінюють один одного на усіх інтервалах даних, а доповнюють, тобто, для кожного окремого інтервалу, потрібно вибрати оптимальний метод.

\section{Метод статистичного самотворення та порівняння методів апроксимації на спостереженнях Z UMa}

З метою порівняння оцінок точності, нами було проведено аналіз методу ''статистичного самотворення''. Іноді його називають калькою з англійської ''бутстреп'' (bootstrap). Сервіс перекладу Google дає `'завантажувач'', що відповідає вживанню терміну в іншому комп'ютерному значенні - при завантаженні операційної системи.

Суть методу полягає в тому, щоб з наявного набору даних сформувати досить велику кількість вибірок, розмір кожної з яких збігається з вихідним інтервалом. Такі вибірки складаються з випадкових комбінацій початкових даних (''самотворення''), а вибір кожного елементу відбувається із можливістью повернення, на відміну від ''перемішування'', при якому усі елементи використовуються один раз у випадковому порядку.

Термін ''bootstrap'' запропонував Ефрон [30], його теорії та застосуванню присвячені монографії, напр. [31-33].

Таким чином, деякі елементи можуть зустрічатися декілька разів, а інші – не зустрічатись взагалі. Для кожної отриманої вибірки визначають значення аналізованих статистичних характеристик з метою вивчити їх розкид, стійкість, розподіл.  Цей метод є вдосконаленою модифікацією метода ''складного ножа'' (jackknife) [30], згідно якого, $n$ нових вибірок даних отримують з початкової видаленням лише однієї точки, послідовно змінюючи номер цієї видаленої точки.

Метод ''bootstrap'' є стандартним для оцінювання точності визначення моментів мінімумів затемнюваних подвійних систем на сайті Секції змінних зір та екзопланет Чеського астрономічного товариства [7] з використанням наближення для повного інтервалу затемнення [34]. Після визначення моменту мінімуму для серії ``самостворених'' рядів даних, визначаються перцентілі 0.025 та 0.975, як границі довірчого інтервалу для ймовірності 95\%. На відміну від стандартного визначення точності, як оцінки середньоквадратичного відхилення $\sigma,$ у [34] задають $\sigma_-$ та $\sigma_+,$ що відповідають приведеним перцентілям.

Використання $\sigma_-$ та/або $\sigma_+$ замість стандартного середньоквадратичного відхилення $\sigma$ призводить до суттєвого зменшення статистичної ваги цієї величини $w={\rm const}/sigma^2.$ У припущенні неперервного нормального (гаусовського) розподілу [35, 36], $\sigma_-=\sigma_+=1.96\sigma$ для цієї ймовірності 95\%. Тобто, зменшення статистичної ваги у $1.96^2\approx3.84\approx4$ рази. Тому потрібно перетворювати два значення на одне, знаходячи середнєквадратичне значення
$\sigma\approx((\sigma_-^2+\sigma_+^2)/2)^{1/2}/1.96;$ значення, що відповідає середній вазі $\sigma\approx((\sigma_-^{-2}+\sigma_+^{-2})/2)^{-1/2}/1.96,$ або інтерперцентильному інтервалу $\sigma\approx(\sigma_-+\sigma_+)/2/1.96.$ При незначном відхиленні $\sigma_-/\sigma_+$ від одиниці, усі три перелічені наближення практично не відрізняються. Суттєве відхилення показує негаусівський характер розподілу похибок, і повинно бути проаналізовано додатково.

Отже, можна рекомендувати використовувати $\sigma$ замість пар $(\sigma_-,\sigma_+).$

Середнєквадратичне значення відхилення для неперервного нормального розподілу, якщо враховувати 95\% середніх значень,  становить $\sigma_{0.95}=0.871\sigma$, отже, при викиданні 5\% 
та обчисленні $\sigma,$ потрібно враховувати даний коефіціент.

Для ілюстрації, з кривої блиску пульсуючої зорі Z UMa був виділений початковий ряд поблизу одного з 94 екстремумів (JD 2455999-2456172, $n=351).$
Спостереження отримані із міжнародної бази даних Французької асоціації спостережень змінних зір (AFOEV) [22].

У якості критерію вибору меж інтервалу, було обрано частину від попереднього до наступного інтервалу, бо ширші інтервали, очевидно, будуть краще апроксимовані сплайном, ніж асимптотичною параболою, де висхідна та низхідна гілки апроксимовані прямими. Квадратичний сплайн враховує можливу кривизну кривої блиску у крайніх підінтервалах та тим самим дозволяє використовувати більші інтервали із більшою кількістю точок.

Було створено 2000 рядів з використанням  генератору випадкових чисел. До них були застосовані ті самі апроксимації, що й для початкового ряду. Апроксимації для початкового ряду з $\pm 1\sigma$ коридором похибок  показані на Рис. 3. Для випадкових рядів показане накладання апроксимацій. Як видно, більшість кривих лежать у порівняно вузькому інтервалі. Але кілька кривих сильно відрізняються. Це є наслідком вибіркової надмірної концентрації даних у окремих точках та відповідного зменшення впливу інших точок.

Для ілюстрації, на рис 3 приведено лише 200 апроксимацій з 2000.
Основна частина апроксимацій є у порівняно вузькому диапазоні.
Апроксимації проведено від початку до кінця згенерованого інтервалу, тому видно, що апроксимації, які суттєво відрізняються від основної смуги, відповідають коротшим інтервалам.

На рис. 4 показана залежність оцінок моментів від номеру (рейтингу по величині) після сортування у порядку збільшення величини, для 2000 ''самостворених'' рядів. Транспонована залежність $i/200$ від $t_e$ називається ''кумулятою''.
Для даної вибірки,  по 2.5\% з обох сторін як раз співпадають із суттєвим надлишком кількості ''викидів'' порівняно із нормальним розподілом, який показано окремою лінією для вибіркового середнього та середньоквадратичного значення відхилення.

Отже, кількість таких ''викидів'' у обидві сторони суттєво менша за 5\%,  тому ''перцентільний'' підхід до оцінки точності $\sigma$ ефективний.

\section{Інші методи аналізу спостережень пульсуючої зорі Z UMa}

У попередніх розділах, були розглянуті методи аналізу даних у інтервалах навколо екстремумів, які не враховують більш віддалені точки.

Втім, найбільш поширеним методом є періодограмний аналіз із застосуванням не ''локальної'', а ''глобальної апроксимації''.
Тест - функція $S(f)=\sigma_C^2/\sigma_O^2=1-\sigma_{O-C}^2/\sigma_O^2=r^2,$ де $\sigma_O^2,$ $\sigma_C^2$ та $\sigma_{O-C}^2$ є, відповідно, дисперсіями даних $''O''$, апроксимованих значень $''C''$, нев'язок (відхилень) $''O-C''$, а $r-$ коефіцієнт кореляції між $''O''$ та $''C''$ (34). Для апроксимації методом найменших квадратів $m_C(t)$, використовується гармонійна функція
\begin{equation}
m_C(t)=C_1-R\cdot\cos(2\pi f\cdot(t-T_0)),
\end{equation}
де $C_1 -$ середнє за період значення (яке, у загальному випадку, не співпадає із вибірковим середнім), $R -$ півам	плітуда, $f=1/P -$ частота, $P -$ період, $T_0 -$ початкова епоха максимуму блиску (тобто, мінімуму зоряної величини $m_C$). На відміну від спрощених алгоритмів, принаймні 6 з яких їх автори називають ''перетворенням Фур'є'' чи ''дискретним перетворенням Фур'є'', у випадку спостережень, нерівномірних у часі, саме метод найменших квадратів є статистично оптимальним та забезпечує незміщеність оцінок параметрів [36, 37, 16].

На рис. 5 показана періодограма у діапазоні частот 0.0001-0.0125 c/d (період от $406^d$ до $10000^d$). Найбільше виділяються чотири піки. Періоди, які їм відповідають, приведені на рисунку. Можна зробити висновок про наявність двох близьких періодів $198.0^d$ та $188.7^d$, а права пара піків відповідають їх гармонікам (подвоєним частотам). Подібна наявність двох близьких періодів ($179.1^d\pm0.3^d$ та $162.4^d\pm0.8^d)$ спостерігалася у іншої системи RX Boo [38]. Огляди зір цього типу приведені, напр. [39-41].

Тому, у програмі MCV, нами було обчислено параметри комбінованої моделі із $s=4$ частотами, дві з яких є подвійними двох перших:
\begin{equation}
m_C(t)= C_1 - \sum_{j=1}^s R_j\cdot\cos(2\pi f_j\cdot(t-T_{0j})).
\end{equation}
Періоди було скореговано методом диференціальних поправок: $P_1=188.88(3),$ $P_2=197.89(4).$ У дужках приведено точність у одиницях останього знаку. Для гармонік, $P_3=P_1/2,$  $P_4=P_2/2.$ Відповідні півамплітуди дорівнюють $0.585^m,$ $0.092^m,$ $0.186^m,$ $0.216^m$ із практично однаковою точністью трохи менше $0.005^m.$ Початкові епохи
2454008.6(3), 2453925.2(8), 2453893.4(8), 2453987.5(3). $C_1=7.898(3).$
Малі оцінки похибок пов'язані із великою загальною кількістью спостережень. Середньоквадратична точність апроксимації у моменти спостережень $\sigma[x_c]=0.010^m.$ Втім, $\sigma[O-C]=0.377^m,$ що значно більше значень $\sigma_{O-C}=0.12^m ... 0.25^m$ (медіанне значення $0.19^m$). Таким чином, індивідуальні цикли пульсацій, хоч і показують цикли биття, мають суттєву аперіодичну складову змінності.

Для довгоперіодичних зір із моноперіодичними та порівняно стабільними кривими блиску, оптимальною апроксимацією є тригонометричний поліном (кінцева сума ряду Фур'є)  із статистично значимим ступенем [42,43]. Для виявлення повільних змін періоду та амплітуди, часто використовують вейвлет-аналіз (wavelet analysis) [44-46], який іноді називають `'спалах-аналіз'' та його модифікацію - ''ковзаючі синуси'' (Running Sine, RS) [47].  Комбінація тригонометричного полінома із локальним профілем затемнення [48-50] теж не ефективна для розглянутого типу коливань, як і визначення ''характерних точок'', як перетину постійного рівня [51-52].

У випадку Z UMa, це не дуже ефективно, тому що ''подвоєння'' частоти пов'язано із ''биттям'' двох коливань із близькими частотами та їх гармоніками.
Тому, більш ефективним буде застосування ''шкалограмного аналізу'' методом ''ковзаючих парабол'' (''running parabola'', RP) [53]. Основна ідея методу полягає у дослідженні систематичних зсувів апроксимації від даних, що дозволяє визначити період та пів-амплітуду для гармонійного сигналу, або їх ефективні значення у випадку квази-періодичних колиівнь (QPO). Із застосуванням $``\Lambda``$ - шкалограмного аналізу [37], було визначено $P_{\Lambda}=190.7^d$ із ефективною півамплітудою $R_{\Lambda}=0.66^m.$
Шкалограми із використанням різних взаємодоповнюючих функцій показані на Рис. 6.
Очікувано, цей метод не дозволяє розрізнити дві близькі частоти, тому період є проміжним, а півамплітуда відповідає сумарним коливанням, а не окремій хвилі.

Ефективна ширина інтервалу згладжування $\Delta t=60^d$ відповідає $\sigma[O-C]=0.22^m.$ Відповідна крива блиску показана на Рис. 7. Добре помітно, що амплітуда коливань при домінуванні швидкої змінності, суттєво занижена. Тому було обрано менше значення $\Delta t=40^d.$ Ця крива показує коливання значно краще.

Також показані локальні апроксимації алгоритмами ''асимптотичної параболи'' (AP) та ''параболічного сплайна'', які були нами реалізовані у програмі MAVKA.
Хоча візуальний вибір інтервалів проводився не лише по індивідуальних точках, а й з врахуванням ''ковзаючих парабол'', використання інтервалів із змінною шириною покращує якість апроксимації.

\section*{Висновки}

Використання взаємодоповнюючих методів апроксимації дозволяє аналізувати сигнали та виявляти різні компоненти періодичної, квазиперіодичної та аперіодичної змінності. Розглянуто різні методи, ефективні для різних типів змінності.

У створеній програмі MAVKA, реалізовано застосування 11 видів апроксимуючих функцій, загалом 21 різна функція, якщо враховувати різні ступені поліномів та симетричних поліномів. Проводяться вдосконалення програми MAVKA та тестування нових методів (включаючи описані у даній статті). У подальшому, сайт програми планується за адресою http://uavso.org.ua/mavka

Використання кусково-заданих функцій дає виграш у часі обчислень порівняно із аналітичними функціями, які потребують обчислення через ряди.

Використання для кусково-заданих функцій інтервалів із межами, що є додатковими параметрами, на відміну від класичних сплайнів, дозволяє зменшити загальну кількість параметрів та покращити якість апроксимації

Реалізовано можливість автоматичного вибору статистично оптимальної апроксимації із обраних користувачем.

Реалізовано можливість автоматичного перебору інтервалів даних із розмічених користувачем.

Досліджено ефективність різних апроксимацій в залежності від ширини інтервалу та розташування його меж. 
Дано рекомендації по розмітці інтервалів.
Показана ефективність локальних наближень із інтерактивно визначеною шириною інтервалу.

Проведено дослідження статистичних властивостей параметрів моделі та апроксимацій методом статистичного самотворення (bootstrap) для тестових та реальних даних. Для додаткової функції ваги, порівняння розподілу 
оцінок коефіціентів із нормальним розподілом проведено [54].

Для реальних спостережень із великими сезонними перервами у спостереженнях, рекомендації можуть бути враховані лише частково.

Досліджено змінність напівправильної пульсуючої подвійної зорі Z UMa.
Показано наявність багатокомпонентної змінності об'єкта, в тому числі, чотирьох періодичних коливань та суттєвої змінності амплітуд та фаз окремих коливань.

Розроблені методи забезпечують статистично оптимальну апроксимацію, що відповідає найкращій точності визначення модельного параметра.

\section*{Подяки}
Дослідження виконано,
 як частина міжнародної програми ``Між-Довготна Астрономія'' (Inter-Longitude Astronomy [55]) та проектів ``Українська віртуальна обсерваторія'' [56,57] та ``Aстроінформатика" [58]. Дякуємо Французській асоціації спостерігачів змінних зір (AFOEV) [22] за спостереження Z UMa та В.І.Марсаковій, Д.Є.Твардовському, Pavol Dubovsky та рецензентам за обговорення.

REFERENCES

%%%
%References should be cited by a numbers in square brackets and listed in the order in which they are cited. %Preparing the reference list please adhere to the following format:
%[1] J. Zinn-Justin, Quantum Field Theory and Critical Phenomena (Oxford University Press, Oxford, 1989).
%[2] P. A. M. Dirac, Rev. Mod. Phys. 21, 392 (1949); https://doi.org/10.1103/RevModPhys.21.392
%[3] B. I. Halperin, T. C. Lubensky, S. Ma, Phys. Rev. Lett. 32, 292 (1974); %https://doi.org/10.1103/PhysRevLett.32.292.
%     for two to five authors.
%[4] A. A. Pevtsov et al., Solar Phys. 289, 593 (2014); https://doi.org/10.1007/s11207-012-0220-5
%     for more than five authors.

\begin{figure}
\includegraphics[width=\textwidth]{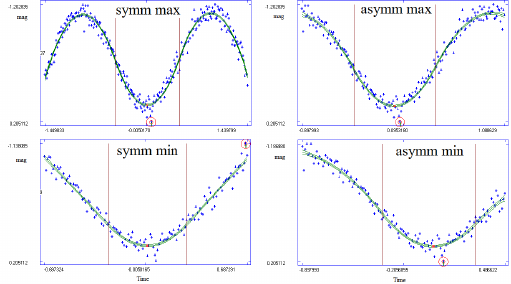}
\caption{Екстремальні за довжиною інтервали для тестування апроксимацій асимптотичною параболою (AP) та параболічним сплайном дефекту 1 (PS) -
Згори - максимальна ширина інтервалу, внизу - найменша. Зліва - для інтервалів із симетричними (symm) даними відносно екстремуму, справа - із суттєво асиметричним (asymm).
Апроксимації проведені параболічним сплайном, крім зліва внизу, де обидві апроксимації практично співпадають. Показано апроксимації $m_C(t)$ та ``коридор похибок'' $m_C(t)\pm\sigma[m_C(t)].$ Діаграми  створені із скріншотів  програми MAVKA, яка показує інформацію про поточну точку (у центрі двох червоних кіл) та дозволяє послідовний перехід між точками або перехід на максимальну або мінімальну точку. На результат апроксимації, номер виділеної точки не впливає.
}\label{fig1}
\end{figure}

\begin{figure}
\includegraphics[width=\textwidth]{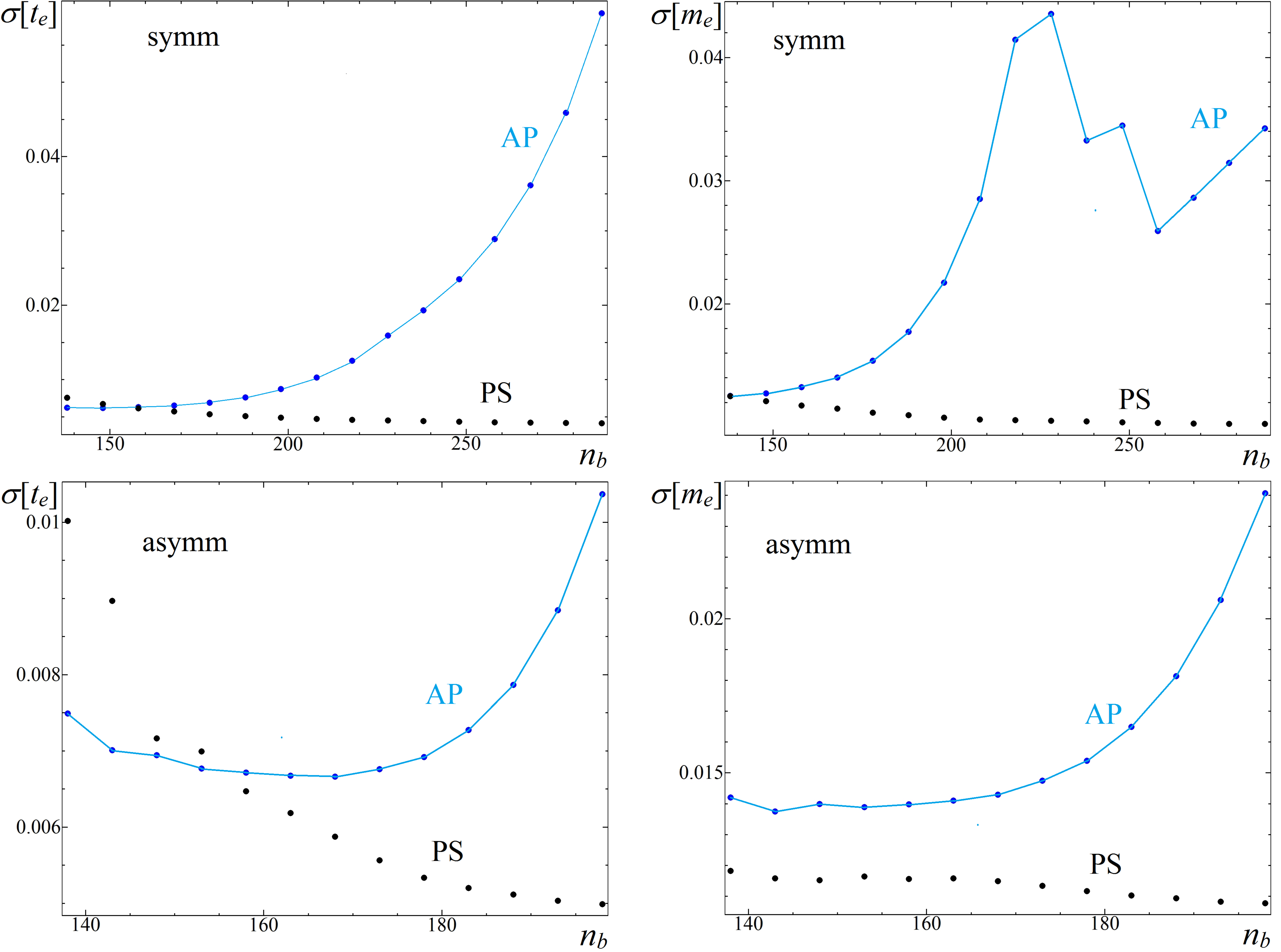}
\caption{Залежність оцінки точності визначення моменту екстремуму $\sigma[t_e]$ та відповідного значення зоряної величини $\sigma[m_e]$ від кількості точок для інтервалів, симетричних (symm) та асиметричних (asymm) відносно екстремума, для асимптотичної параболи (AP) та параболічного сплайну (PS).
}\label{fig2}
\end{figure}

\begin{figure}
\includegraphics[width=0.45\textwidth]{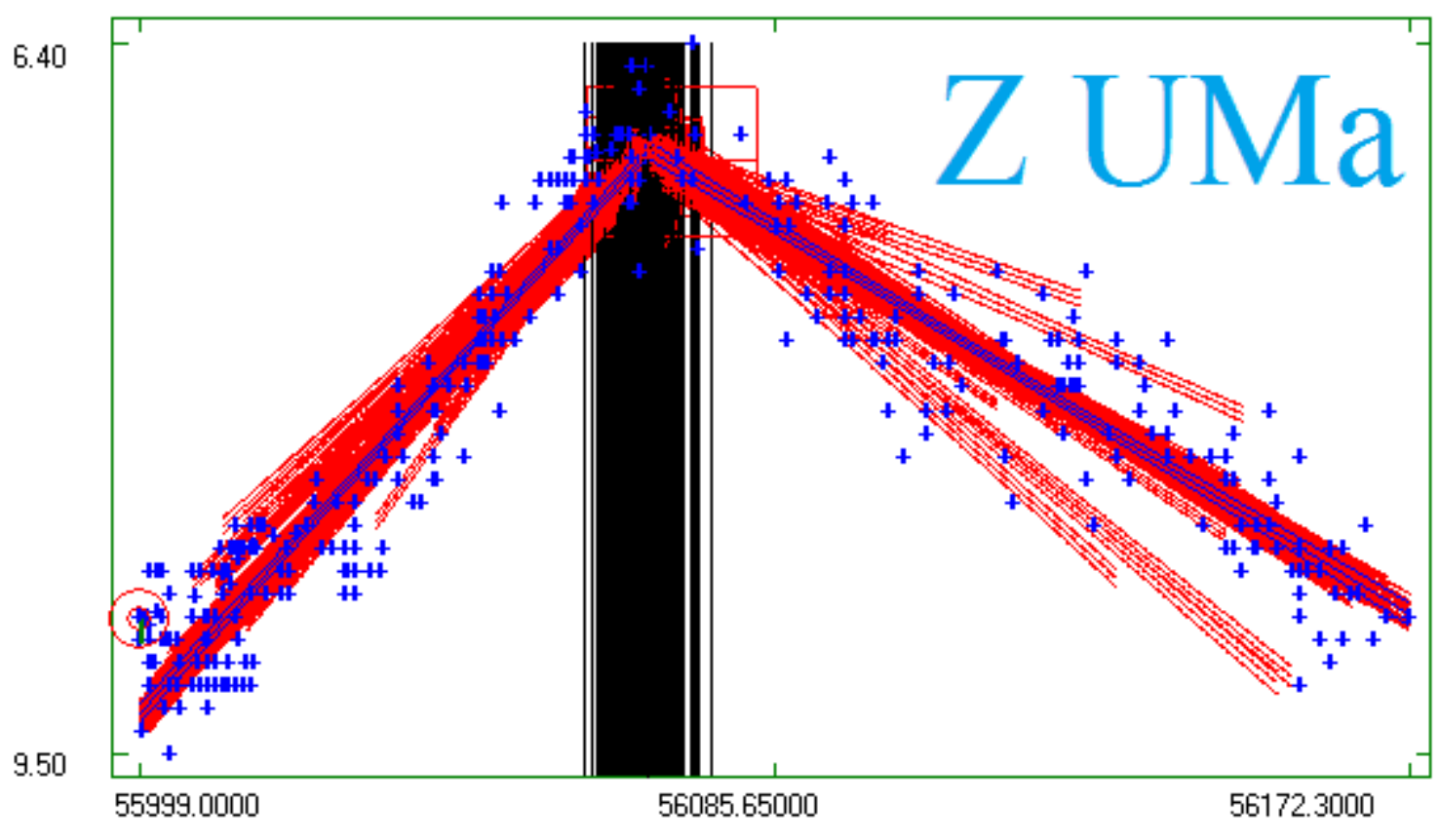}
\caption{Частина кривої блиску напівправильної пульсуючої змінної Z UMa біля одного з екстремумів. Апроксимації $(m_C(t))$ асимптотичною параболою показані разом із коридором похибок $(m_C(t)\pm\sigma[m_C(t)])$ для 200 рядів методу статистичного самотворення (bootstrap). Відповідні криві проведені від початку до кінця створеного ряду. Сині лінії всередині червоної смуги демонструють апроксимацію та коридор похибок для початкового набору даних.
}\label{fig3}
\end{figure}

\begin{figure}
\includegraphics[width=0.45\textwidth]{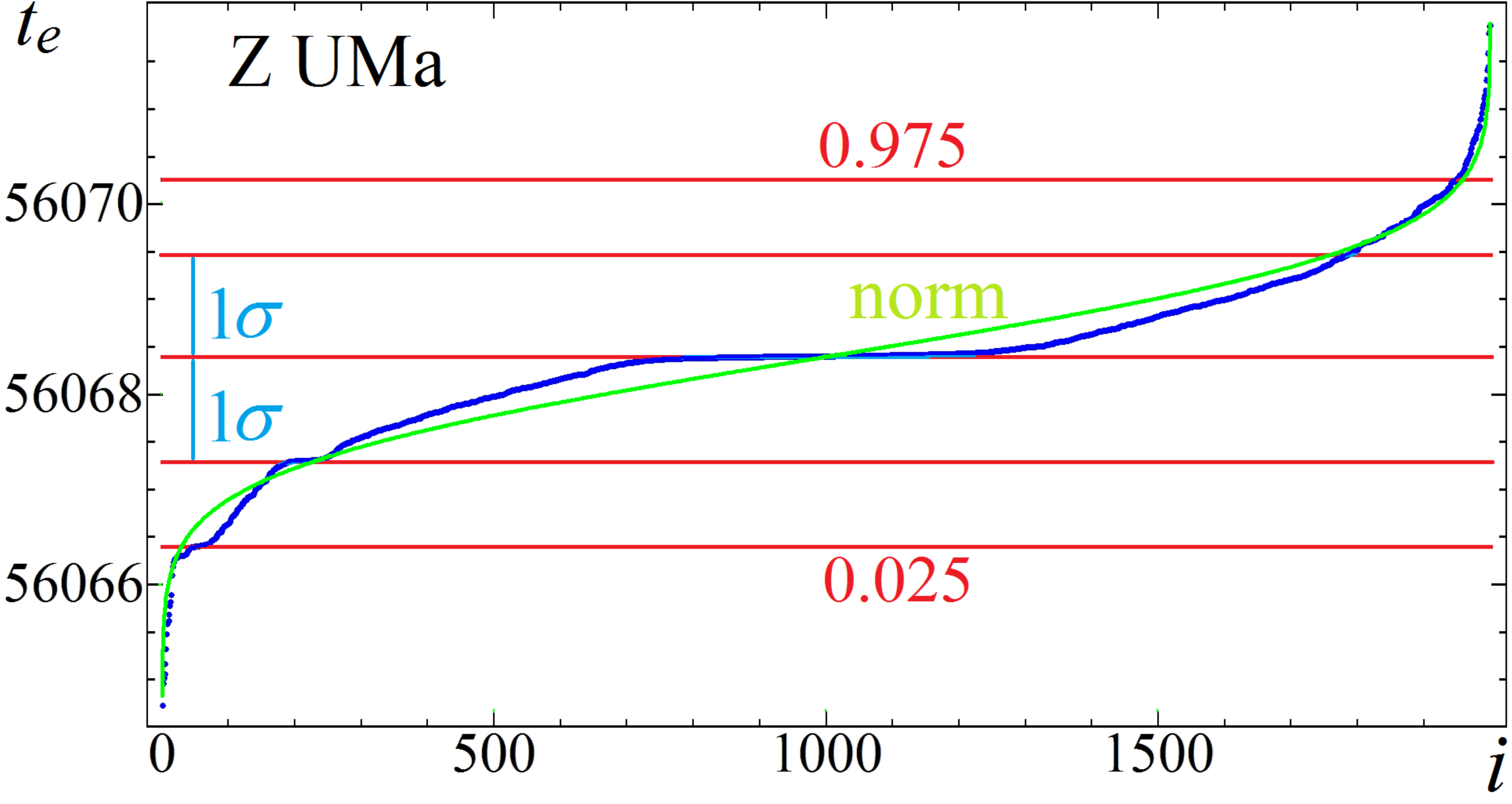}
\caption{Залежність оцінки моменту екстремуму $t_e$ від номеру реалізації після сортування по величині. У методі статистичного самотворення, не враховуються перші та останні 2.5\% чисел (показані верхньою та ніжньою лініями). Внутрішніми горизонтальними лініями показане значення для початкового ряду та коридор похибок. Окремо наведена теоретична крива для нормального розподілу із відповідними вибірковими значеннями.
}\label{fig4}
\end{figure}

\begin{figure}
\includegraphics[width=0.45\textwidth]{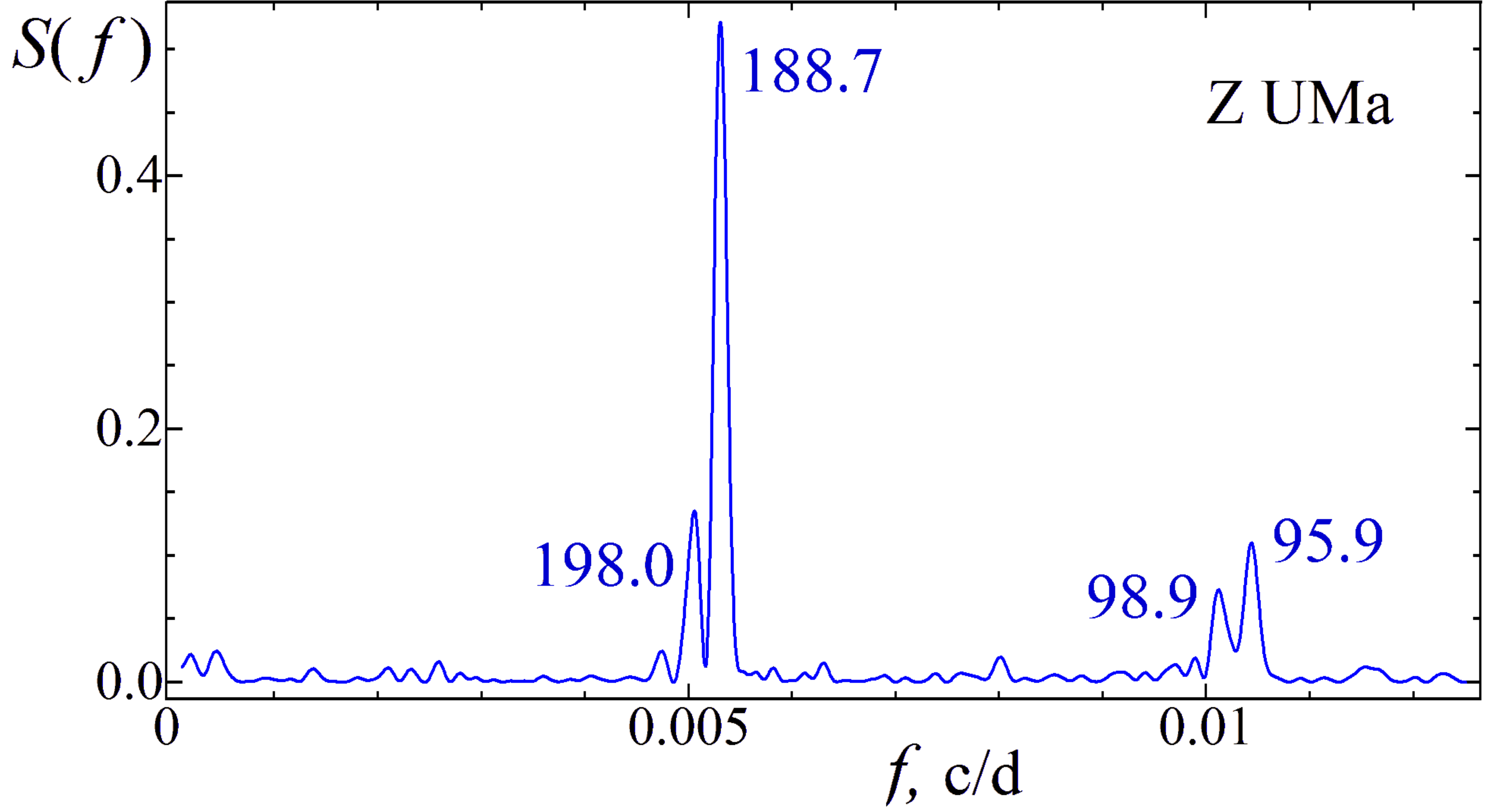}
\caption{
Періодограма $S(f)$ для спостережень Z UMa з бази даних AFOEV. Статистично значущі піки позначаються значеннями відповідних періодів.
%Periodogram $S(f)$ for the observations of Z UMa from the AFOEV database. The statistically significant peaks are marked by values of the corresponding periods.
}\label{fig5}
\end{figure}

\begin{figure}
\includegraphics[width=\textwidth]{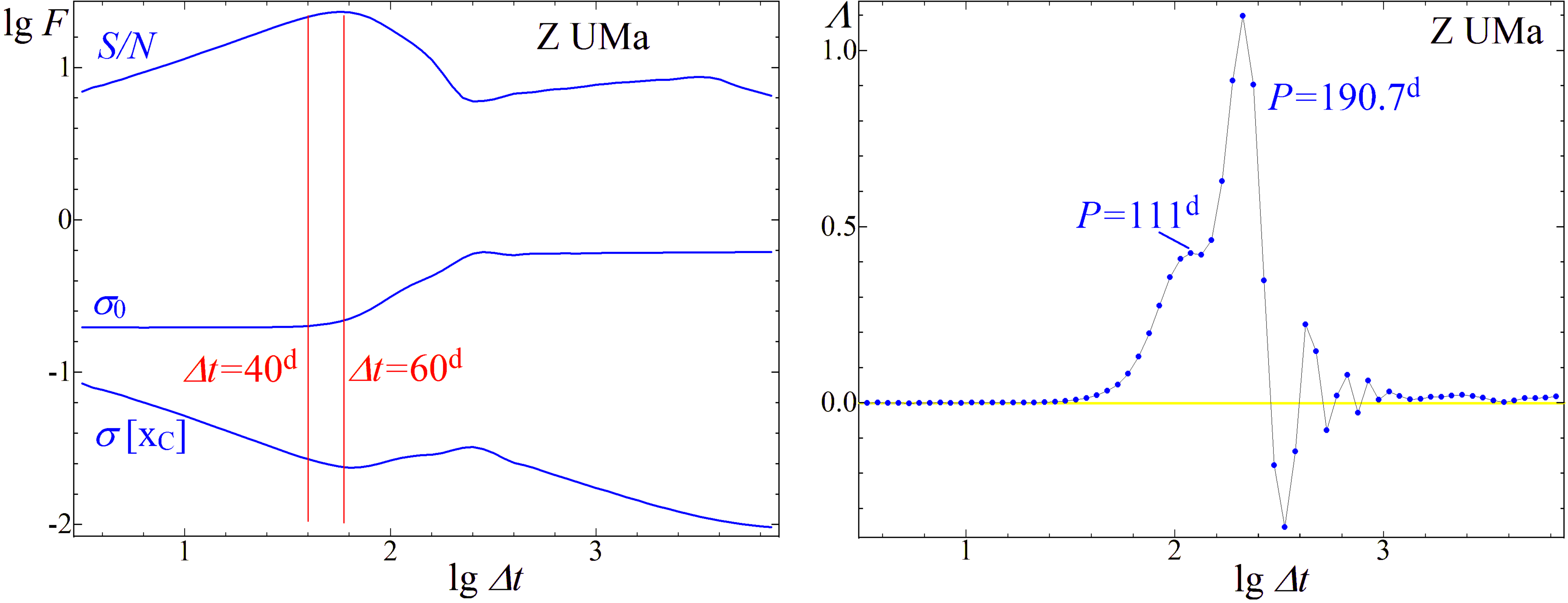}
\caption{Зліва:  шкалограма ``ковзаючою параболою'' [53] для Z UMa. Вертикальні лінії відповідають статистично оптимальним значенням $\Delta t=60^d$ і прийнятому значенню $\Delta t = 40^d $, що відповідає рідкому видимому подвоєнню частоти. Праворуч: $''\Lambda - ''$ шкалограма [15] та відповідні оцінки періоду.
%Left: ``Running parabola'' scalegrams [] for Z UMa. Vertical lines correspond to statistically optimal value $\Delta t=60^d$ and the adopted value  $\Delta t=40^d$ corresponding to  occasional apparent doubling of the frequency. Right: $''\Lambda-''$ scalegram [29] and the corresponding period estimates.
}\label{fig6}
\end{figure}

\begin{figure}
\includegraphics[width=\textwidth]{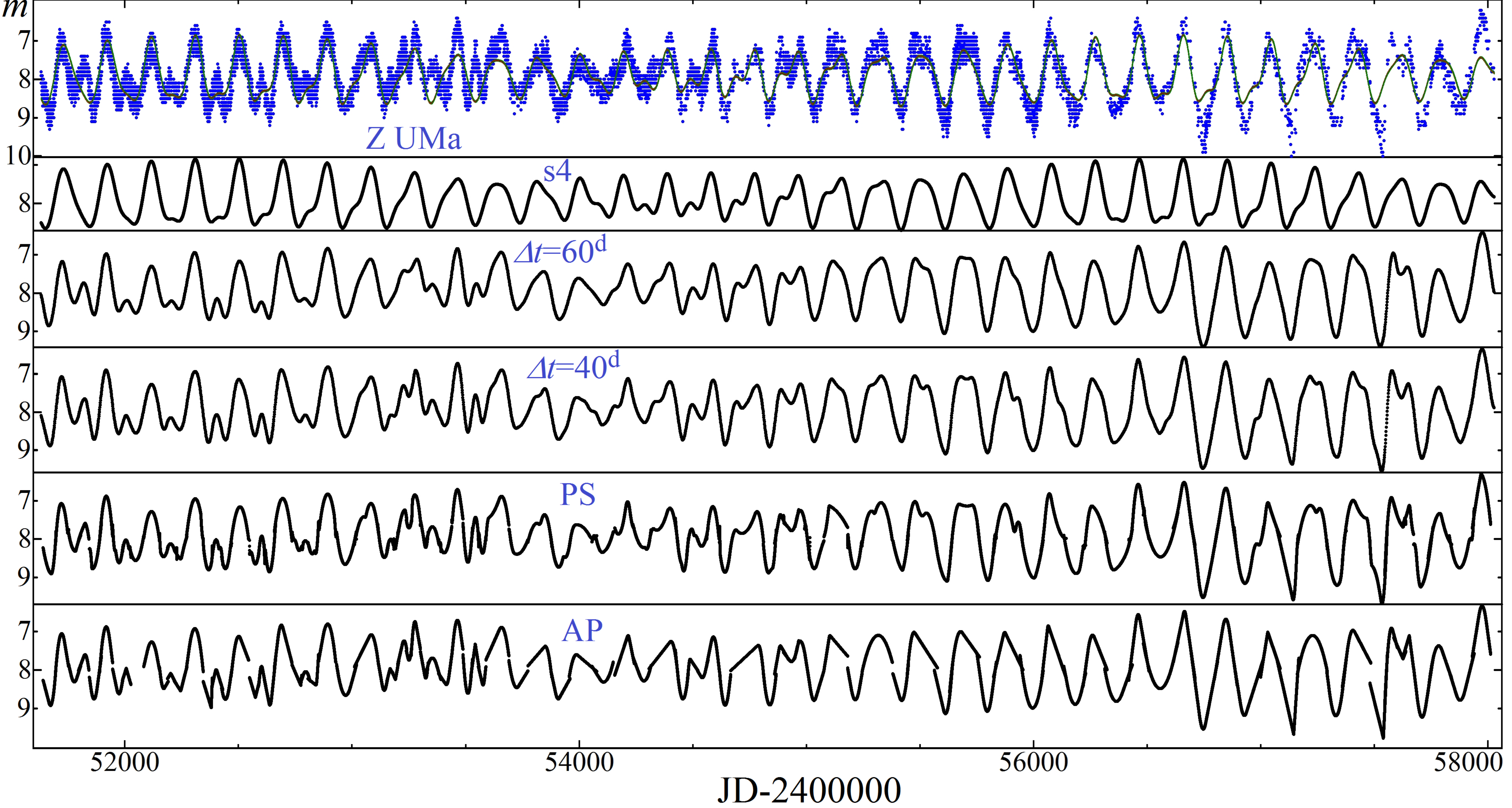}
\caption{Спостереження Z UMa та їх наближення: s4 - наближення з 4 синусоїдальними компонентами; $\Delta t=60^d$ та $\Delta t=40^d-$ апроксимація методом ''ковзаючих парабол'' із відповідною шириною фільтра; PS -  апроксимація параболічним сплайном по розмічених інтервалах поблизу екстремумів; AP -  апроксимація ''асимптотичною параболою'' тих самих інтервалів.
%Observations of Z UMa and their approximations: s4 - approximation with 4 sinusoidal components; $\Delta t=60^d$ and $\Delta t=40^d$ -
}\label{fig7}
\end{figure}

\end{document}